\documentclass[conference,10pt]{IEEEtran}

\usepackage[T1]{fontenc}
\usepackage[utf8]{inputenc}
\usepackage{microtype}
\usepackage{graphicx}
\usepackage{booktabs}
\usepackage{array}
\usepackage{tabularx}
\usepackage{listings}
\usepackage{fancyvrb}
\usepackage{xcolor}
\usepackage{tikz}
\usetikzlibrary{positioning, arrows.meta, shapes.geometric, fit, calc,
                backgrounds, decorations.pathreplacing}
\usepackage{pgfplots}
\pgfplotsset{compat=1.18}
\usepackage{amsmath}
\usepackage{url}
\usepackage[hidelinks]{hyperref}
\usepackage{cleveref}

\definecolor{lstbg}{gray}{0.97}
\definecolor{lstkw}{rgb}{0.10,0.30,0.65}
\definecolor{lstcm}{rgb}{0.30,0.55,0.30}
\definecolor{lststr}{rgb}{0.65,0.10,0.10}

\lstdefinestyle{cstyle}{
  language=C,
  basicstyle=\ttfamily\footnotesize,
  keywordstyle=\color{lstkw}\bfseries,
  commentstyle=\color{lstcm}\itshape,
  stringstyle=\color{lststr},
  backgroundcolor=\color{lstbg},
  frame=single,
  rulecolor=\color{lstbg},
  breaklines=true,
  showstringspaces=false,
  columns=fullflexible,
  numbers=none,
  xleftmargin=0pt,
  xrightmargin=0pt,
  aboveskip=4pt,
  belowskip=4pt,
}
\lstdefinestyle{shellstyle}{
  basicstyle=\ttfamily\scriptsize,
  backgroundcolor=\color{lstbg},
  frame=single,
  rulecolor=\color{lstbg},
  breaklines=true,
  columns=fullflexible,
  numbers=none,
  xleftmargin=0pt,
  xrightmargin=0pt,
  aboveskip=4pt,
  belowskip=4pt,
}
\title{Inference Pipelines as Operating-System Objects:\\
Priority Scheduling and Constant-Footprint Streaming 
for Microcontroller Neural Inference}

\author{\IEEEauthorblockN{Dimitrios Kafetzis}
\IEEEauthorblockA{SynapticOS Project, Hamburg, Germany}}

\begin{document}
\maketitle

\begin{abstract}
Microcontroller runtimes that host neural-network inference treat the
inference pipeline --- pre-processing, accelerator invocation,
post-processing --- as application code: every project re-implements
stage sequencing, intermediate-buffer sizing, and completion
signalling around a library call. We argue these are
operating-system concerns, and present the Phase~2 inference engine
of SynapticOS, an open-source runtime built on Zephyr, which makes
the pipeline itself a first-class OS object. A pipeline is drawn
from a static pool, validated at build time against a canonical
stage ordering, and executed by a priority job scheduler
(\textsc{realtime} $>$ \textsc{normal} $>$ \textsc{best-effort},
FIFO within class) with per-job completion semaphores, cancellation,
and a bounded job table --- no heap allocation anywhere on the
inference path. Stage buffers are planned size-aware: exact
capacities are computed from stage configuration and runtime tensor
geometry for the nine built-in processors, with a bounded
$4\times$ worst-case fallback for user-supplied stages. Because
every intermediate lives in an ephemeral arena region that is reset
per frame, streaming workloads run at constant memory footprint.
We evaluate on the NXP FRDM-MCXN947 (Cortex-M33 at 150~MHz) and on
the \texttt{qemu\_cortex\_m3} continuous-integration target, with
the model stage executing a deterministic stub NPU kernel on both
--- honest engine-overhead baselines, not silicon throughput. On
the board, the full scheduler path costs 92~\textmu s of wall time
over the Phase~1 direct-HAL bracket (1{,}130 vs.\
1{,}038~\textmu s), of which dispatch inside the profiled window is
1~\textmu s; a 30-frame face-detection pipeline (resize, normalize,
quantize, model, decode, non-maximum suppression) averages
4.63~ms per frame (215.8~FPS, stub model stage included) against
31.1~ms under QEMU soft-float, at a constant 2{,}784-byte arena
peak that returns to zero after every frame with zero fragmentation
by construction. The MCXN947's PowerQuad DSP is routed and
self-calibrated for FFT and Q15 matrix multiply; measured end-to-end
speedups over the software kernels are 5.51$\times$ (256-point FFT)
and 1.66$\times$ ($16{\times}16$ matmul) with all wrapper costs
included --- short of the phase plan's $\geq$10$\times$ target,
which we report as missed and analyse rather than re-scope.
Stage-level profiling fires at pipeline stage boundaries, closing a
known Phase~1 gap, live on the board. The engine adds 3.8~KB of
flash to the QEMU build and 20.7~KB to the FRDM build (shell,
PowerQuad routing, and calibration/bench harness included). A
99-test suite across 13 ZTEST suites passes 100\% under emulation.
SynapticOS is released under Apache~2.0 at
\url{https://github.com/Dimitrios-Kafetzis/SynapticOS}.

\end{abstract}


\section{Introduction}
\label{sec:intro}

The previous paper in this project~\cite{synapticos-p1} made the case
that a microcontroller with an on-die neural processing unit (NPU)
deserves an operating system that treats inference as a first-class
workload, and delivered the foundation for one: a tensor-aware bump
allocator with persistent and ephemeral lifetimes, a four-state NPU
hardware-abstraction layer with a deterministic software stub, a
model-lifecycle registry, and a cycle-accurate profiling surface, all
running on Zephyr~\cite{zephyr} on the NXP
FRDM-MCXN947~\cite{nxp-mcxn947}. That foundation deliberately stopped
below the inference engine: it could allocate a tensor and invoke an
accelerator, but the code that turns a camera frame into a
classification --- resize, normalize, quantize, invoke, decode,
suppress --- remained, as it does on every production MCU stack
today, application code.

This paper is about moving that code into the operating system. On
the dominant stacks --- Zephyr or FreeRTOS hosting TensorFlow Lite
Micro~\cite{zephyr,freertos,tflm} --- an inference pipeline exists
only as a convention in the application: the developer calls the
pre-processing routines in the right order, sizes each intermediate
buffer by hand against the worst case, invokes the interpreter, and
arranges completion signalling and prioritisation with raw RTOS
primitives if more than one model or more than one client is
involved. Each of these steps is a recurring, structurally identical
problem, and each has a failure mode that is discovered late:
mis-ordered stages produce silently wrong tensors, undersized
intermediates corrupt memory or fail at the worst moment, and ad-hoc
completion signalling is where concurrency bugs live
(\cref{sec:discussion:race} documents one we found in our own
engine).

\subsection{Pipelines as OS Objects}
\label{sec:intro:position}

Our position is that the inference pipeline has the same claim to OS
citizenship that files, sockets, and threads have: it is a recurring
workload structure with invariants that the system can check and
resources that the system can plan. Concretely, three observations
drive the Phase~2 design:

\begin{enumerate}
\item \textbf{Pipelines have a checkable shape.} Real inference
pipelines are a chain: zero or more pre-processing stages, exactly
one model invocation, zero or more post-processing stages. An OS
object can enforce that ordering at construction time and validate
completeness before first use, converting a class of silent
data-corruption bugs into immediate \texttt{-EINVAL} returns.
\item \textbf{Stage buffers are plannable.} For built-in processing
stages, the exact output size is a closed-form function of the stage
configuration and the runtime input geometry ($12{\times}12{\times}3$
bytes for a resize to $12{\times}12$ over 3 channels; $4\times$ the
element count for a uint8-to-float32 normalize). The engine can
therefore allocate exact-fit intermediates from the ephemeral arena
and reserve a bounded worst case only for stages it cannot see
inside.
\item \textbf{Inference completion is a scheduling event, not a
callback convention.} Once jobs from multiple clients target one
accelerator, someone must decide dispatch order, expose completion,
and define cancellation. Doing this once, in the OS, with a bounded
job table and per-job semaphores, is cheaper and safer than every
application growing its own queue.
\end{enumerate}

\subsection{Contributions}
\label{sec:intro:contributions}

Building on the Phase~1 memory, HAL, registry, and profiling
subsystems~\cite{synapticos-p1}, this paper contributes:

\begin{enumerate}
\item A \textbf{pipeline-as-OS-object abstraction}
(\cref{sec:pipeline}): a construction API over a static pipeline
pool with canonical stage ordering enforced at add-time, build-time
validation, and a worst-case memory estimate computed before first
execution.
\item \textbf{Size-aware stage-buffer planning}
(\cref{sec:pipeline:buffers}): exact output capacities computed from
stage configuration plus runtime tensor geometry for the nine
built-in processors, with a bounded $4\times$ fallback (64-byte
floor) for custom stages, all served from the Phase~1 ephemeral
arena.
\item A \textbf{priority job scheduler for inference on a single MCU
core} (\cref{sec:sched}): three priority classes with FIFO ordering
within a class, a dedicated scheduler thread, per-job completion
semaphores, cancellation semantics, and a bounded job table --- no
heap allocation on the submission or completion path.
\item \textbf{Constant-footprint streaming}
(\cref{sec:streaming}): per-frame ephemeral-arena reset gives zero
fragmentation across unbounded frame counts; the face-detection
demo sustains 30 frames at a constant 2{,}784-byte arena peak ---
measured on the board --- that returns to zero after every frame.
\item \textbf{Live stage-level profiling} (\cref{sec:profiling}):
the Phase~1 four-mark profiler is now driven at pipeline stage
boundaries, closing the ``\texttt{syn prof last} returns no data''
gap reported in the Phase~1 paper; when profiling is disabled at
runtime the marks reduce to four early-return branch tests per
inference.
\item An \textbf{honest-baseline evaluation} (\cref{sec:eval})
continuing the Phase~1 methodology: all timing runs through the
deterministic stub NPU are labeled as engine-overhead baselines, the
engine's own cost is isolated from model latency, and the one phase
acceptance criterion the measurements did not meet --- the
$\geq$10$\times$ PowerQuad speedup target, measured at 5.51$\times$
and 1.66$\times$ with wrapper costs included --- is reported as
missed and analysed (\cref{sec:eval:pq,sec:discussion:limits})
rather than re-scoped.
\end{enumerate}

The engine is roughly 1{,}700 lines of new C (pipeline and scheduler
core, nine processors, and the shared software DSP kernels) behind
the frozen Phase~1 public headers plus one new header
(\texttt{syn\_process.h}) for stage configuration structures.

\subsection{Scope}
\label{sec:intro:scope}

This paper reports the Phase~2 engine as validated on the
\texttt{qemu\_cortex\_m3} continuous-integration target and live on
the FRDM-MCXN947 board (transcripts captured 2026-07-12; released
as v0.2.0). Three boundaries matter for interpretation. First, the
model stage executes the deterministic stub NPU kernel inherited
from Phase~1~\cite{synapticos-p1}, not the eIQ Neutron
silicon~\cite{nxp-neutron}; every latency number is an
engine-overhead baseline, not a throughput claim. The PowerQuad DSP
numbers, by contrast, are real hardware. Second, the PowerQuad
measurement came in below the phase plan's acceptance target, and
we treat that result as a finding to analyse
(\cref{sec:eval:pq}), not a blank to defer. Third, two scheduler
parameters (\texttt{deadline\_us}, \texttt{preemptible}) are
accepted and recorded but not yet acted on: deadline-aware dispatch
and layer-granularity preemption are Phase~3 work
(\cref{sec:discussion:limits}).

\subsection{Paper Organisation}
\label{sec:intro:org}

\Cref{sec:related} positions the pipeline and scheduler against
existing MCU inference runtimes. \Cref{sec:pipeline} presents the
pipeline object and its size-aware buffer planning;
\cref{sec:sched} the priority scheduler; \cref{sec:processors} the
built-in processors and software DSP kernels;
\cref{sec:streaming-profiling} constant-footprint streaming and the
profiling wire-up. \Cref{sec:eval} evaluates footprint, latency,
streaming behaviour, and test coverage; \cref{sec:discussion}
discusses a scheduler race caught by the test suite, limitations,
and the roadmap. \Cref{sec:conclusion} concludes.

\section{Related Work}
\label{sec:related}

The Phase~1 paper~\cite{synapticos-p1} surveyed the OS-level
landscape --- RTOS hosts for embedded AI, tensor memory managers,
and accelerator HALs --- and we do not repeat that survey here. This
section focuses on the two questions Phase~2 answers: who owns the
\emph{pipeline}, and who owns the \emph{dispatch decision}, in
existing MCU inference stacks.

\subsection{Pipeline Ownership}
\label{sec:related:pipeline}

TensorFlow Lite Micro~\cite{tflm} owns the graph \emph{inside} the
model boundary: operators within a \texttt{.tflite} flatbuffer are
sequenced by the interpreter, and their intermediates are planned in
the interpreter's arena. Everything outside that boundary ---
resizing and normalising the camera frame, quantising to the input
scale, decoding output tensors to boxes, non-maximum suppression ---
is application code with hand-sized buffers. The same split holds
for ExecuTorch~\cite{executorch}, whose ahead-of-time-planned
runtime is similarly scoped to the model graph, and for
$\mu$TVM~\cite{microtvm}, which compiles the graph to straight-line
C with statically planned tensors but leaves pre- and
post-processing to the caller. CMSIS-NN and
CMSIS-DSP~\cite{cmsis-nn,cmsis-dsp} sit one layer lower still:
libraries of kernels that callers wire together by hand, with
caller-supplied buffers and no runtime notion of a pipeline at all.

SynapticOS inverts the boundary: the OS object spans the
\emph{whole} pipeline, from raw sensor tensor to application-level
result, with the model invocation as one stage among several. This
is what lets the engine do end-to-end buffer planning
(\cref{sec:pipeline:buffers}) --- the intermediates that TFLM cannot
see, because they live outside the flatbuffer, are exactly the ones
our built-in stages describe in closed form. The two designs are
complementary rather than competing: a TFLM interpreter invocation
is a natural implementation of our model stage, and remains the
planned integration path (\cref{sec:discussion:roadmap}).

\subsection{Dispatch Ownership}
\label{sec:related:dispatch}

TFLM's interpreter is single-threaded and synchronous by design; the
project's own guidance for concurrent workloads is to serialise
access externally or instantiate one interpreter per thread. In
practice, production Zephyr and FreeRTOS deployments wrap inference
in one of two idioms: submit closures to a system work queue, or
dedicate a thread per model and mediate with queues and semaphores.
Both idioms re-derive, per application, the questions a scheduler
answers once: in what order do competing requests run, how is
completion exposed, and what does cancellation mean.

The work-queue idiom is the closest structural relative of our
scheduler, and its known sharp edges motivated our departures from
it. A Zephyr work queue executes work items strictly FIFO at a
single thread priority --- there is no notion that a wake-word
inference matters more than a background telemetry classification.
Prioritisation requires multiple queues at different thread
priorities, at one thread stack each, and cancellation of an
already-running item is undefined. Our scheduler keeps the single
dedicated thread (one stack, one context) but adds a priority-aware
pick over a bounded job table (\textsc{realtime} $>$
\textsc{normal} $>$ \textsc{best-effort}, FIFO within class),
per-job completion semaphores, and defined cancel semantics
(\cref{sec:sched:jobs}). Classical real-time theory offers the next
steps up this ladder --- deadline-driven ordering is well understood
since Liu and Layland~\cite{liu73sched} --- and the job parameters
already carry \texttt{deadline\_us} and \texttt{preemptible} fields
for exactly that evolution, but we deliberately ship the simpler
priority-class scheduler first and report the fields as inert
(\cref{sec:discussion:limits}).

On application processors, ONNX Runtime~\cite{onnxrt} and similar
runtimes do own scheduling across execution providers, with
thread-pool parallelism and inter-op concurrency; as with their
memory and dispatch machinery, the footprint and threading
assumptions do not transfer to a Cortex-M budget. MLPerf
Tiny~\cite{banbury21mlperftiny} measures exactly the class of
workload we target but is silent on scheduling: its harness runs one
model in isolation, which is precisely the situation where
pipeline-and-scheduler machinery looks unnecessary --- until a
second model, or a second client, arrives.

\subsection{Streaming Memory Behaviour}
\label{sec:related:streaming}

Region- and arena-based memory management is long-established
folklore~\cite{regionalloc}, and the Phase~1
paper~\cite{synapticos-p1} positioned the SynapticOS arena against
TFLM's interpreter-owned arena in detail. Phase~2 adds the
\emph{streaming} claim: because every pipeline intermediate is an
ephemeral-arena tensor and the application resets the ephemeral
region at frame boundaries, per-frame memory behaviour is identical
across unbounded frame counts --- the sawtooth in
\cref{fig:sawtooth} rather than the monotonic creep or
fragmentation-driven failure that a general-purpose heap invites.
TFLM achieves a comparable steady state within one interpreter
invocation by re-planning the same arena offsets each invoke; our
version extends the property across the whole pipeline, including
the stages TFLM does not see, and makes the reset an explicit,
application-visible lifecycle event with an observable statistic
(\texttt{syn mem stats} reporting the peak and the return to zero).

\section{The Pipeline Object}
\label{sec:pipeline}

A SynapticOS pipeline is an ordered chain of stages --- zero or more
pre-processors, exactly one model invocation, zero or more
post-processors --- owned and validated by the runtime.
\Cref{lst:pipeapi} reproduces the construction API from the frozen
Phase~1 header \texttt{syn\_infer.h}; the Phase~2 work reported here
is the implementation behind it.

\begin{lstlisting}[style=cstyle,caption={Pipeline construction API
(\texttt{include/synaptic/syn\_infer.h}, frozen).},label={lst:pipeapi},
captionpos=b]
syn_pipeline_t *syn_pipeline_create(const char *name);
int  syn_pipeline_add_preprocess(syn_pipeline_t *pipe,
                                 syn_preprocess_fn_t fn,
                                 void *config);
int  syn_pipeline_add_model(syn_pipeline_t *pipe,
                            syn_model_handle_t model);
int  syn_pipeline_add_postprocess(syn_pipeline_t *pipe,
                                  syn_postprocess_fn_t fn,
                                  void *config);
int  syn_pipeline_build(syn_pipeline_t *pipe);
void syn_pipeline_destroy(syn_pipeline_t *pipe);
\end{lstlisting}

\subsection{Static Pool and Construction Invariants}
\label{sec:pipeline:construction}

Pipelines are drawn from a static pool of four slots --- consistent
with the no-heap discipline of the Phase~1
allocator~\cite{synapticos-p1}, nothing on the inference path ever
touches a general-purpose heap. \texttt{syn\_pipeline\_create()}
claims a slot under the engine mutex and returns \texttt{NULL} on
pool exhaustion. Every subsequent handle crossing the API boundary
is validated by range (the pointer must lie inside the pool) and by
liveness (the slot's \texttt{in\_use} flag), so a stale or forged
handle fails fast instead of corrupting engine state.

Two invariants are enforced at \emph{add-time} rather than at build
time, so the failing call itself returns the error:

\begin{itemize}
\item \textbf{Canonical ordering.} A pre-processor added after the
model stage, or a post-processor added before it, is rejected with
\texttt{-EINVAL} and a log line naming the pipeline. The accepted
grammar is exactly
$\mathit{pre}^{*}\;\mathit{model}\;\mathit{post}^{*}$.
\item \textbf{Single model stage.} A second
\texttt{syn\_pipeline\_add\_model()} returns \texttt{-EALREADY};
multi-model graphs are out of scope for this phase
(\cref{sec:discussion:limits}). The model handle itself is checked
against the Phase~1 registry at add-time, so a dangling handle is
caught before the pipeline can be built.
\end{itemize}

Stage capacity is bounded by
\texttt{CONFIG\_SYNAPTIC\_\allowbreak MAX\_\allowbreak PIPELINE\_\allowbreak STAGES}
(default 8, range 4--16), and a built pipeline is immutable: further stage additions
return \texttt{-EPERM} until the pipeline is destroyed. Destruction
returns the slot to the pool after cancelling any jobs still queued
against the pipeline (each such job completes with
\texttt{-ECANCELED} through the normal completion path, so no waiter
deadlocks on a destroyed pipeline).

\texttt{syn\_pipeline\_build()} validates completeness (a model
stage must be present), then walks the stage chain to compute a
worst-case memory estimate: the model's declared SRAM requirement
plus the worst-case capacity of every intermediate buffer the chain
will allocate (using the $4\times$ rule of
\cref{sec:pipeline:buffers} and the model's declared output size).
The estimate is logged at build time --- the
\texttt{hello\_inference} transcript in \cref{sec:eval:latency}
shows \texttt{est.\ 4106 bytes} for a single-stage pipeline over a
model declaring 4{,}096~B of SRAM and a 10-byte output --- and gives
the application a pre-execution answer to ``will this pipeline fit
the arena?'' that today's hand-wired stacks can only discover by
running out of memory.

\subsection{Size-Aware Stage-Buffer Planning}
\label{sec:pipeline:buffers}

Every stage writes its output into a fresh ephemeral-arena tensor
allocated by the engine immediately before the stage runs. The
planning question is the capacity of that tensor, and the engine
answers it at two levels of knowledge.

\textbf{Exact capacities for built-in stages.} When the stage
function is one of the nine built-in processors
(\cref{sec:processors}), the engine computes the exact output size
from the stage configuration and the \emph{runtime} geometry of the
incoming tensor. \Cref{tab:capacities} lists the closed forms. The
resize capacity, for example, is $w \cdot h \cdot c$ bytes with
$w{\times}h$ from the stage config and the channel count $c$ read
from the incoming tensor's last dimension at execution time --- the
plan adapts to the actual input rather than a declared worst case.

\begin{table}[t]
\caption{Exact output capacities for built-in stages, computed from
stage config and runtime input geometry ($n$ = input payload bytes).}
\label{tab:capacities}
\centering
\renewcommand{\arraystretch}{1.2}
\setlength{\tabcolsep}{4pt}
\begin{tabular}{@{}ll@{}}
\toprule
\textbf{Stage} & \textbf{Output capacity (bytes)} \\
\midrule
\texttt{image\_resize}    & $w \cdot h \cdot c$ (config $w,h$; input channels $c$) \\
\texttt{image\_normalize} & $n \cdot 4$ (uint8/int8 $\rightarrow$ float32) \\
\texttt{quantize\_int8}   & $n / 4$ (float32 $\rightarrow$ int8) \\
\texttt{audio\_mfcc}      & $\lfloor n/4F \rfloor \cdot C \cdot 4$ (frame len $F$, coeffs $C$) \\
\midrule
\texttt{softmax}          & $n$ if float32 input, else $n \cdot 4$ \\
\texttt{argmax}           & 8 (one \texttt{syn\_classification\_t}) \\
\texttt{top\_k}           & $k \cdot 8$ (config $k$) \\
\texttt{nms}              & $n$ (kept boxes $\subseteq$ candidates) \\
\texttt{dequantize}       & $n \cdot 4$ (int8 $\rightarrow$ float32) \\
\bottomrule
\end{tabular}
\end{table}

\textbf{Bounded fallback for custom stages.} A stage function the
engine does not recognise gets
$\max(4n, 64)$ bytes, where $n$ is the input payload size. The
$4\times$ factor covers the worst legal expansion in the type system
(a 1-byte-per-element tensor promoted to float32); the 64-byte floor
keeps degenerate inputs (an argmax result feeding a custom stage)
from receiving unusably small buffers. The fallback is deliberately
a \emph{bound}, not a guess: a custom stage that needs more than
$4\times$ must claim its own memory, and the engine's estimate stays
conservative rather than optimistic.

\textbf{The model-feed guarantee.} Whichever rule produced the
capacity, the stage that feeds the model stage is additionally
raised to the model's declared input size, so a pre-processing chain
can never hand the accelerator a short buffer. In the
face-detection pipeline this rule is a no-op (the quantize output is
exactly the model input size); it exists for chains whose final
pre-processor legitimately shrinks data below the declared input.

\textbf{The stage-buffer convention.} Stage functions receive the
planned capacity in \texttt{out->size} and must set the final
geometry --- \texttt{shape}, \texttt{ndim}, \texttt{dtype},
\texttt{size} --- before returning. This convention does double
duty: the capacity check inside each built-in stage converts a
planning bug into \texttt{-ENOMEM} at the offending stage (with a
log line naming the shortfall) rather than a buffer overrun, and the
final geometry lets the next stage plan against actual rather than
worst-case dimensions. Stage outputs, including the pipeline's final
result, live in the ephemeral arena region and remain valid until
the application calls \texttt{syn\_mem\_reset\_ephemeral()} --- the
lifecycle contract that \cref{sec:streaming} builds streaming on.

\begin{figure*}[t]
\centering
\begin{tikzpicture}[
  font=\small,
  stage/.style={
    rectangle, draw=black!70, thick, rounded corners=1pt,
    minimum height=11mm, text width=17mm, align=center, inner sep=2pt,
  },
  buf/.style={font=\scriptsize\ttfamily, text=black!65},
  appside/.style={rectangle, draw=black!50, dashed, thick,
    minimum height=11mm, text width=17mm, align=center, inner sep=2pt},
  arrow/.style={-Latex, semithick, black!75},
  caplbl/.style={font=\scriptsize, text=black!60, align=center},
]

\node[appside]                     (input)  at (0,0)
  {frame\\[-1pt]{\scriptsize $24{\times}24{\times}3$ u8}};
\node[stage, fill=blue!10]         (resize) [right=7.5mm of input]
  {resize\\[-1pt]{\scriptsize $\rightarrow 12{\times}12{\times}3$}};
\node[stage, fill=blue!10]         (norm)   [right=7.5mm of resize]
  {normalize\\[-1pt]{\scriptsize u8 $\rightarrow$ f32}};
\node[stage, fill=blue!10]         (quant)  [right=7.5mm of norm]
  {quantize\\[-1pt]{\scriptsize f32 $\rightarrow$ i8}};
\node[stage, fill=orange!20]       (model)  [right=7.5mm of quant]
  {model\\[-1pt]{\scriptsize stub NPU}};
\node[appside]                     (decode) [right=7.5mm of model]
  {decode\\[-1pt]{\scriptsize app code}};
\node[stage, fill=green!12]        (nms)    [right=7.5mm of decode]
  {NMS\\[-1pt]{\scriptsize built-in}};

\draw[arrow] (input)  -- node[caplbl, above=1pt] {1728 B}  (resize);
\draw[arrow] (resize) -- node[caplbl, above=1pt] {432 B}   (norm);
\draw[arrow] (norm)   -- node[caplbl, above=1pt] {1728 B}  (quant);
\draw[arrow] (quant)  -- node[caplbl, above=1pt] {432 B}   (model);
\draw[arrow] (model)  -- node[caplbl, above=1pt] {64 B}    (decode);
\draw[arrow] (decode) -- node[caplbl, above=1pt] {boxes}   (nms);

\begin{scope}[on background layer]
\node[rectangle, draw=black!40, thick, rounded corners=2pt,
      fill=black!3, inner sep=3.5mm,
      fit=(resize)(norm)(quant)(model),
      label={[font=\footnotesize\itshape, text=black!55]below:%
        pipeline object (stages sequenced and buffers planned by the engine;
        intermediates in the ephemeral arena)}] {};
\end{scope}

\end{tikzpicture}
\caption{The \texttt{face\_detection} pipeline as built. Solid boxes
are engine-owned stages; dashed boxes are application code (the frame
source and the model-specific box decoding, after which the
application invokes the built-in NMS directly). Arrow labels give the
exact stage-buffer capacities planned from stage configuration and
runtime tensor geometry (\cref{tab:capacities}); the model output is
raised to the 64-byte floor. All intermediates are ephemeral-arena
tensors reclaimed by the per-frame reset.}
\label{fig:pipeline}
\end{figure*}
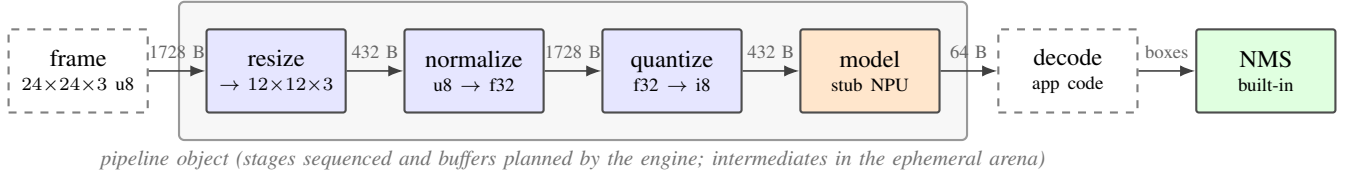

\Cref{fig:pipeline} shows the face-detection pipeline as built,
with the planned capacity of each intermediate.

\section{The Priority Job Scheduler}
\label{sec:sched}

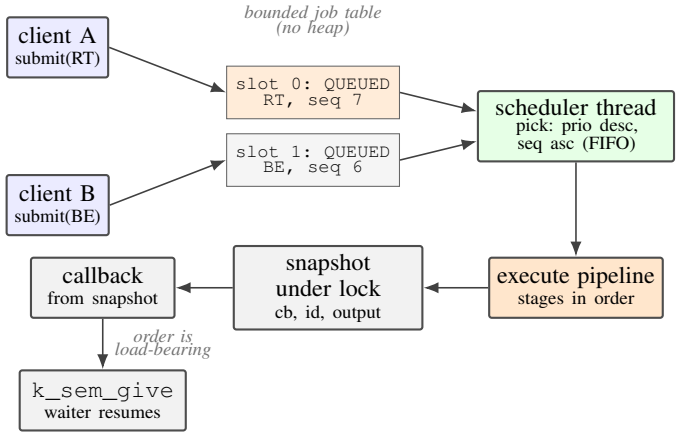
\begin{figure}[t]
\centering
\resizebox{\columnwidth}{!}{%
\begin{tikzpicture}[
  font=\small,
  box/.style={rectangle, draw=black!70, thick, rounded corners=1pt,
    align=center, inner sep=3pt, minimum height=8mm},
  slot/.style={rectangle, draw=black!60, thin, align=center,
    minimum width=17mm, minimum height=7mm, font=\scriptsize\ttfamily},
  arrow/.style={-Latex, semithick, black!75},
  note/.style={font=\scriptsize\itshape, text=black!55, align=center},
]

\node[box, fill=blue!8]  (c1) at (0,2.1)  {client A\\[-2pt]{\scriptsize submit(RT)}};
\node[box, fill=blue!8]  (c2) at (0,0)    {client B\\[-2pt]{\scriptsize submit(BE)}};

\node[slot, fill=orange!15] (s1) at (3.4,1.5)  {slot 0: QUEUED\\[-2pt]RT, seq 7};
\node[slot, fill=black!4]   (s2) at (3.4,0.6)  {slot 1: QUEUED\\[-2pt]BE, seq 6};
\node[note] (tbl) at (3.4,2.45) {bounded job table\\[-2pt](no heap)};

\node[box, fill=green!10, text width=24mm] (sched) at (6.9,1.05)
  {scheduler thread\\[-2pt]{\scriptsize pick: prio desc,\\[-3pt] seq asc (FIFO)}};

\node[box, fill=orange!20, text width=21mm] (exec) at (6.9,-1.1)
  {execute pipeline\\[-2pt]{\scriptsize stages in order}};
\node[box, fill=black!5, text width=23mm] (snap) at (3.6,-1.1)
  {snapshot under lock\\[-2pt]{\scriptsize cb, id, output}};
\node[box, fill=black!5, text width=17mm] (cb) at (0.6,-1.1)
  {callback\\[-2pt]{\scriptsize from snapshot}};
\node[box, fill=black!5, text width=20mm] (sem) at (0.6,-2.6)
  {\texttt{k\_sem\_give}\\[-2pt]{\scriptsize waiter resumes}};

\draw[arrow] (c1.east) -- (s1.west);
\draw[arrow] (c2.east) -- (s2.west);
\draw[arrow] (s1.east) -- ($(sched.west)+(0,0.2)$);
\draw[arrow] (s2.east) -- ($(sched.west)+(0,-0.2)$);
\draw[arrow] (sched) -- (exec);
\draw[arrow] (exec) -- (snap);
\draw[arrow] (snap) -- (cb);
\draw[arrow] (cb) -- node[note, right=1pt] {order is\\[-3pt]load-bearing} (sem);

\end{tikzpicture}%
}
\caption{Scheduler architecture. Submissions claim slots in a bounded
job table; a dedicated thread picks the highest-priority queued job
(FIFO within class) and executes its pipeline non-preemptively.
Completion snapshots the callback context under the engine mutex,
fires the callback from the snapshot, and gives the per-job semaphore
\emph{last} --- the ordering whose violation
\cref{sec:discussion:race} dissects.}
\label{fig:scheduler}
\end{figure}

Execution is decoupled from construction: a built pipeline is
submitted as a \emph{job}, and a dedicated scheduler thread decides
what runs next. \Cref{fig:scheduler} shows the architecture;
\cref{lst:jobapi} the submission API.

\begin{lstlisting}[style=cstyle,caption={Job submission API and
parameters (\texttt{syn\_infer.h}, frozen).},label={lst:jobapi},
captionpos=b]
typedef struct {
    syn_priority_t  priority;    /* RT > NORMAL > BE  */
    uint32_t        deadline_us; /* recorded; Phase 3 */
    bool            preemptible; /* recorded; Phase 3 */
    syn_infer_cb_t  callback;
    void           *user_data;
} syn_infer_params_t;

syn_job_id_t syn_infer_submit(syn_pipeline_t *pipe,
                              const syn_tensor_t *input,
                              const syn_infer_params_t *params);
int syn_infer_wait(syn_job_id_t job, uint32_t timeout_ms);
int syn_infer_cancel(syn_job_id_t job);
int syn_infer_get_result(syn_job_id_t job, syn_tensor_t *output);
\end{lstlisting}

\subsection{Bounded Job Table}
\label{sec:sched:jobs}

Jobs live in a fixed table of
\texttt{CONFIG\_SYNAPTIC\_\allowbreak MAX\_\allowbreak CONCURRENT\_\allowbreak JOBS}
slots (default 2, range 1--4). A slot carries the job's state machine
(\textsc{free} $\rightarrow$ \textsc{queued} $\rightarrow$
\textsc{running} $\rightarrow$ \textsc{done}/\textsc{error}/%
\textsc{cancelled} $\rightarrow$ \textsc{free}), the submission
parameters, a monotonic sequence number, the result tensor
descriptor, and a per-job completion semaphore. Submission takes the
engine mutex, checks the count of \textsc{queued}-plus-%
\textsc{running} jobs against a runtime-adjustable concurrency cap
(\texttt{syn\_infer\_set\_max\_concurrent()}), claims a free slot,
and wakes the scheduler --- constant work, no allocation. Job IDs
are 32-bit, monotonically assigned, and skip the reserved invalid
value on wrap-around; lookups match IDs only against non-free slots,
so a stale ID for a recycled slot returns \texttt{-ENOENT} rather
than aliasing the new occupant.

A completed job's slot is \emph{held} until the client consumes the
outcome with \texttt{syn\_infer\_get\_result()}, which returns the
output descriptor (for \textsc{done}), the error (for
\textsc{error}), or \texttt{-ECANCELED}, and only then frees the
slot. This makes result delivery reliable at the cost that a client
who abandons a job leaks a slot until the table's bound is felt ---
a deliberate trade documented with the API, and the reason
\texttt{syn\_infer\_run\_sync()} consumes the slot even on timeout.

\subsection{Dispatch Policy}
\label{sec:sched:policy}

The scheduler thread blocks on a wake semaphore, then selects the
\textsc{queued} job with the highest priority class, breaking ties
by lowest sequence number: strict priority across classes
(\textsc{realtime} $>$ \textsc{normal} $>$ \textsc{best-effort}),
FIFO within a class. With a job table of at most four slots the
$O(\textit{slots})$ linear scan is cheaper than any queue structure
worth maintaining.

Dispatch is \emph{non-preemptive}: priority governs which job starts
next, and a \textsc{realtime} arrival overtakes any queued
\textsc{best-effort} job, but it does not interrupt one that is
already running. The worst-case priority inversion is therefore one
pipeline execution, which is acceptable at Phase-2 job granularity
and is precisely what the recorded-but-inert \texttt{preemptible}
flag and the layer-granularity preemption work in Phase~3 are
scoped to fix (\cref{sec:discussion:limits}). For the same reason,
``concurrent'' jobs are concurrent in submission but serialised in
execution: one scheduler thread executes one pipeline at a time on
the single Cortex-M33 core, and the concurrency cap bounds queue
depth, not parallelism.

The scheduler runs in a dedicated 2~KB-stack thread at preemptible
Zephyr priority 8, rather than on the system work queue. \Cref{sec:related:dispatch} gives the comparative
rationale; the operational reasons are that inference latency should
not inherit head-of-line blocking from unrelated system work items,
and that a single well-known thread makes the engine's CPU share
visible and tunable through the standard Zephyr thread analyser
rather than smeared across work-queue callbacks.

\subsection{Completion Protocol}
\label{sec:sched:completion}

Completion crosses from the scheduler thread to the client through
two channels: an optional callback and the per-job semaphore, in
that order. After the pipeline returns, the scheduler takes the
engine mutex, records the result, moves the job to
\textsc{done}/\textsc{error}, and --- critically --- snapshots the
callback pointer, user data, job ID, and output descriptor into
locals before releasing the mutex. The callback then fires from the
snapshot, and only afterwards is the semaphore given.

The ordering is load-bearing. The semaphore give is the moment a
blocked waiter can resume, consume the result, and resubmit --- at
which point the slot may be reinitialised for a new job. Firing the
callback before the give, and from a snapshot rather than from the
slot, guarantees the callback observes the job it belongs to.
Our first implementation gave the semaphore first and read the slot
afterwards, and the test suite caught the resulting slot-reuse race
deterministically; \cref{sec:discussion:race} reconstructs that bug
in full as a case study.

Cancellation is defined by job state: a \textsc{queued} job is
marked \textsc{cancelled} and its semaphore given (waiters observe
\texttt{-ECANCELED}); a \textsc{running} job returns
\texttt{-EBUSY} (nothing interrupts an executing pipeline); a
finished job returns \texttt{-EALREADY}. Destroying a pipeline
cancels its queued jobs through the same path
(\cref{sec:pipeline:construction}).

\subsection{Synchronous Convenience Path}
\label{sec:sched:runsync}

\texttt{syn\_infer\_run\_sync()} packages the common
one-model-no-stages case: it creates a transient pipeline (named
\texttt{run\_sync} in the engine's build-time log line), builds it,
submits at the caller's priority, waits with a 5-second bound,
copies the result into a caller buffer if one is supplied (or hands
back the arena descriptor if not), and destroys the pipeline. It is
the path the \texttt{hello\_inference} sample and the
\texttt{syn infer run} shell command use, and --- because it
includes pipeline create, build, and destroy in every call --- it
is also the worst case for engine overhead, which is exactly why
\cref{sec:eval:latency} measures it.

\section{Built-in Processors and DSP Kernels}
\label{sec:processors}

The size-aware planning of \cref{sec:pipeline:buffers} works because
the engine ships the stages that real pipelines are made of. Phase~2
provides nine built-in processors --- four pre-processors and five
post-processors --- plus the two software DSP kernels
(FFT and Q15 matrix--vector multiply) that were declared in the
Phase~1 DSP HAL but returned \texttt{-ENOTSUP} until
now~\cite{synapticos-p1}.

\subsection{The Nine Built-in Stages}
\label{sec:processors:builtin}

All nine stages implement the uniform stage signature of
\cref{lst:pipeapi} and honour the stage-buffer convention (capacity
in, final geometry out). Configuration travels through small
per-stage structs in the new public header \texttt{syn\_process.h}
--- the one header added in Phase~2; the Phase~1 headers remain
frozen.

\textbf{Pre-processors.}
\texttt{image\_resize} performs edge-aligned bilinear interpolation
(pixel centres mapped, borders clamped) on uint8/int8 images of up
to four channels. \texttt{image\_normalize} applies per-channel
$(x - \mu_c)/\sigma_c$, producing float32.
\texttt{quantize\_int8} computes
$q = \mathrm{round}(x / s) + z$ with saturation to the int8 range.
\texttt{audio\_mfcc} computes MFCC features~\cite{davis80mfcc} as
Hamming window $\rightarrow$ FFT (through the DSP HAL)
$\rightarrow$ power spectrum $\rightarrow$ triangular mel filterbank
$\rightarrow$ log $\rightarrow$ DCT-II, with the complex FFT working
buffer drawn from the Phase~1 scratch pool so it is reclaimed at the
same lifecycle boundary as every other intermediate. The MFCC
implementation makes four deliberate, documented simplifications
relative to a librosa-parity frontend: non-overlapping frames, no
pre-emphasis filter, natural (not decadic) log, and an unnormalized
DCT-II. It targets keyword-spotting models trained with the same
frontend; models trained against librosa features need the Phase-3
parity pass (\cref{sec:discussion:limits}).

\textbf{Post-processors.}
\texttt{softmax} accepts int8 logits (dequantized in place via an
optional config) or float32 and produces float32 probabilities
through the DSP HAL's numerically-stable softmax.
\texttt{argmax} emits a single
\texttt{syn\_classification\_t} record; \texttt{top\_k} emits $k$ of
them by repeated selection ($O(kn)$, no sort buffer, no
allocation). \texttt{nms} implements greedy per-class non-maximum
suppression over packed \texttt{syn\_bbox\_t} records (24~B each:
two corners, score, class), dropping boxes under a score threshold
and suppressing overlaps above an IoU threshold, capped at a
configured maximum. \texttt{dequantize} inverts the int8
quantization to float32.

One boundary is deliberate: converting raw model output tensors to
\texttt{syn\_bbox\_t} candidates (anchor or grid decoding) is
application code, not a built-in --- the decode step encodes
model-family knowledge (anchor layouts, cell geometry) that does not
generalise, whereas NMS over decoded boxes does. The face-detection
sample's \texttt{decode\_cells()} plays this role in
\cref{sec:streaming}.

\subsection{Software DSP Kernels and the PowerQuad Path}
\label{sec:processors:dsp}

Phase~2 implements the two outstanding DSP HAL entry points as
shared software kernels in \texttt{syn\_dsp\_soft.c}:

\begin{itemize}
\item \texttt{fft\_f32}: an iterative radix-2 Cooley--Tukey
FFT~\cite{cooley65fft} over interleaved complex float32, in-place
capable, with bit-reversal permutation and lengths restricted to
powers of two in $[2, 1024]$.
\item \texttt{mat\_mult\_q15}: a saturating Q15 matrix--vector
multiply with 64-bit accumulation, arithmetic shift down by 15, and
saturation to int16 --- the CMSIS-DSP-style
contract~\cite{cmsis-dsp} for fixed-point classifier heads.
\end{itemize}

The kernels are \emph{shared}, not stub-only: the QEMU stub DSP
backend calls them directly, and the MCXN947 backend routes the same
entry points to the PowerQuad engines
(\texttt{PQ\_TransformCFFT}, \texttt{PQ\_MatrixMultiplication}),
retaining the software kernels as the reference implementation and
fallback. The software kernels are validated by the FFT unit suite
(DC, impulse, in-place, and argument-validation cases among them;
\cref{sec:eval:tests}) and by the Phase~1 cross-check suite pattern;
the hardware path is then validated \emph{against them} by a
boot-time self-calibration pass on the board: the FFT path confirms
the PowerQuad's $1/N$ output-gain model on a known input (measured
gain 1024/16384 at $N{=}16$), and the matmul path runs a Q15
known-answer and saturation check, before the backend advertises
hardware routing. Two Phase~1 tests that asserted \texttt{-ENOTSUP}
for these entry points were updated to assert success --- the only
Phase~1 test-suite changes in this phase. Measured on the board with
all wrapper costs included, the hardware paths deliver 5.51$\times$
(256-point FFT) and 1.66$\times$ ($16{\times}16$ Q15 matmul) over
the software kernels --- real speedups, but below the phase plan's
$\geq$10$\times$ acceptance target; \cref{sec:eval:pq} reports the
measurement and its decomposition, and
\cref{sec:discussion:limits} the honest reading.

\section{Constant-Footprint Streaming and Profiling}
\label{sec:streaming-profiling}

\subsection{Constant-Footprint Streaming}
\label{sec:streaming}

The pipeline engine allocates every intermediate --- and the final
result --- from the ephemeral region of the Phase~1
arena~\cite{synapticos-p1}, and never frees anything individually.
The contract with the application is a single lifecycle event: when
the result of frame $n$ has been consumed, the application calls
\texttt{syn\_mem\_reset\_ephemeral()}, which reclaims all of frame
$n$'s intermediates (and the scratch pool) in constant time. Frame
$n{+}1$ then allocates into the identical addresses under the
identical plan.

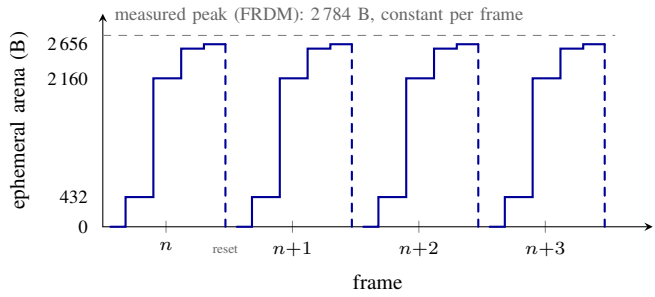
\begin{figure}[t]
\centering
\begin{tikzpicture}
\begin{axis}[
  width=\columnwidth, height=44mm,
  xlabel={frame}, ylabel={ephemeral arena (B)},
  xmin=0, xmax=4.35, ymin=0, ymax=3100,
  xtick={0.5,1.5,2.5,3.5}, xticklabels={$n$,$n{+}1$,$n{+}2$,$n{+}3$},
  ytick={0,432,2160,2656},
  yticklabels={0,432,{2\,160},{2\,656}},
  ytick style={draw=none},
  yticklabel style={font=\scriptsize},
  xticklabel style={font=\scriptsize},
  label style={font=\footnotesize},
  axis lines=left, line cap=round,
  clip=false,
]
\pgfplotsinvokeforeach{0,1,2,3}{
  \addplot[thick, blue!60!black, const plot, forget plot] coordinates {
    (#1.06,0) (#1.18,432) (#1.40,2160) (#1.62,2592)
    (#1.80,2656) (#1.97,2656)
  };
  \draw[thick, blue!60!black, dashed]
    (axis cs:#1.97,2656) -- (axis cs:#1.97,0);
}
\draw[black!50, thin, dashed] (axis cs:0,2784) -- (axis cs:4.05,2784);
\node[font=\scriptsize, text=black!60, anchor=south west]
  at (axis cs:0.02,2810)
  {measured peak (FRDM): 2\,784 B, constant per frame};
\node[font=\tiny, text=black!60, anchor=north] at (axis cs:0.97,-130)
  {reset};
\end{axis}
\end{tikzpicture}
\caption{Constant-footprint streaming in \texttt{face\_detection}
(schematic; payload values from the buffer plan of
\cref{fig:pipeline}, descriptor overhead omitted). Each frame
allocates the same staircase of stage buffers --- 432~B resize,
1{,}728~B normalize, 432~B quantize, 64~B model output --- and the
per-frame \texttt{syn\_mem\_reset\_ephemeral()} (dashed drops,
marked \emph{reset}) returns occupancy to zero, so frame
$n{+}k$ is bit-identical in memory behaviour to frame $n$ for any
$k$. The FRDM run (\cref{sec:eval:facedet}) measures the peak at
2{,}784~B --- the payload staircase plus descriptor overhead ---
returning to zero on all 30 frames (120 allocations, 30 resets);
QEMU's integer-KB display reports the same peak as 2~KB.}
\label{fig:sawtooth}
\end{figure}

The consequence is a memory profile that is a flat sawtooth
(\cref{fig:sawtooth}): per-frame peak occupancy is a constant of the
pipeline, not a function of frame count, and fragmentation is zero
by construction --- there is no free-list to fragment. Where a
general-purpose heap under a streaming inference workload degrades
with uptime (the classical checkerboard pattern the Phase~1 paper
measured the arena against), the arena-backed pipeline's 30th,
300th, and 3-millionth frames are bit-identical in memory behaviour
to the first. The claim is validated, not just argued: the
face-detection run in \cref{sec:eval:facedet} reports its arena
peak and its return to zero after every one of 30 frames through
\texttt{syn mem stats}.

The \texttt{face\_detection} sample is the streaming acceptance
workload for the phase, exercising every subsystem this paper
describes. A deterministic synthetic frame source (a dark gradient
plus a bright $8{\times}8$ blob that advances one grid cell per
frame) stands in for the OV7670 camera so the demo runs identically
under QEMU and on the board. Each $24{\times}24{\times}3$ uint8
frame flows through a pipeline built once and reused for all
frames --- resize to $12{\times}12{\times}3$, normalize to float32,
quantize to int8, model --- submitted at \textsc{realtime} priority;
the application then decodes the model's per-cell scores to
candidate boxes (\texttt{decode\_cells()}, the app-side stand-in for
anchor decoding; \cref{sec:processors:builtin}) and calls the
built-in NMS directly. The frame closes with the ephemeral reset.
Camera (DVP) and LCD overlay integration are tracked hardware
bring-up items; the pipeline, scheduler, profiling, and
post-processing path in the sample is the production path.

\subsection{Profiling the Live Inference Path}
\label{sec:profiling}

The Phase~1 paper shipped a four-mark profiling API
(start, preprocess-done, NPU-done, end) and reported, as a known
limitation, that nothing invoked it: \texttt{syn prof last}
returned \texttt{"No profiling data available"} on real
hardware~\cite{synapticos-p1}. Phase~2 closes that gap by firing
the marks at pipeline stage boundaries inside the engine itself:
start before the first stage, preprocess-done at the model-stage
entry, NPU-done at model-stage exit, end after the last
post-processor. A pipeline with no model stage cannot arise (build
validation), and pre- or post-free pipelines degenerate cleanly ---
the corresponding interval is simply empty. Because the engine owns
stage sequencing, every inference through the scheduler is
attributed automatically; applications add no instrumentation code.

From the four marks the profiler derives per-stage times, total
time, the arena high-water mark, and an NPU utilisation ratio
(NPU interval over total). Utilisation is exactly the engine-overhead
metric the honest-baseline methodology wants: on the stub backend it
reads 99\% (\cref{sec:eval:latency}), meaning the engine's dispatch
and buffer planning consume 1\% of the profiled window --- and on
real Neutron silicon it will expose, rather than hide, any engine
overhead that model acceleration reveals.

Two properties are worth stating precisely. First, profiling is
runtime-switched (\texttt{syn prof enable}/\texttt{disable}); when
disabled, each mark is an early-return branch test --- four branch
tests per inference, not zero cost, but no timestamp reads, no
memory-statistics calls, and no state writes. Second, the profiler
keeps the \emph{last} completed result in a single slot; since the
scheduler serialises execution (\cref{sec:sched:policy}), marks from
different jobs cannot interleave, but a client that profiles a
specific job should read \texttt{syn\_prof\_get\_last()} before the
next job completes. The new \texttt{syn infer run <model>} shell
command runs a named model through the full scheduler path and
prints the profile, making the closed Phase~1 gap directly visible
on hardware. \Cref{lst:frdmshell} reproduces the live board
session: where the Phase~1 paper's transcript showed
\texttt{"No profiling data available"} at this point, the same
command now returns the stage breakdown of the inference that just
ran --- 1{,}069~\textmu s profiled, of which the stub model stage
is 1{,}068~\textmu s, leaving roughly 1~\textmu s of engine
dispatch on real silicon (\cref{sec:eval:latency}).

\begin{lstlisting}[style=shellstyle,float=tp,caption={FRDM-MCXN947
shell session (captured 2026-07-12): a live inference through the
scheduler and its profile. ANSI escapes and one interleaved
asynchronous log line elided; text otherwise verbatim
(\texttt{community/phase2/serial-frdm-shell.log}).},
label={lst:frdmshell},captionpos=b]
uart:~$ syn infer run test_classify
Model 'test_classify': class 0 (confidence 127), 1130 us
Use 'syn prof last' for the stage breakdown.
uart:~$ syn prof last
Last inference:
  Total:       1069 us
  Preprocess:  1 us
  NPU:         1068 us
  Postprocess: 1 us
  Memory peak: 1792 bytes
uart:~$ syn mem stats
Arena: 0/114688 bytes (peak 1792)
Scratch: 0/16384 bytes
Allocations: 6, Resets: 2
uart:~$ syn npu state
NPU state: IDLE
\end{lstlisting}

\section{Evaluation}
\label{sec:eval}

We evaluate the Phase~2 engine with the methodology of the Phase~1
paper~\cite{synapticos-p1}: every number that passes through the
deterministic stub NPU backend is labeled an engine-overhead
baseline rather than silicon throughput, the engine's own cost is
isolated from the (stub) model latency, and an acceptance criterion
that was not met is reported as missed
(\cref{sec:eval:pq}). QEMU runs use \texttt{icount shift=6}, so
their timings are deterministic emulated time; QEMU numbers were
captured on 2026-07-11 (\texttt{community/phase2/results-qemu.md}).
FRDM-MCXN947 numbers were captured live on the board on 2026-07-12
from the v0.2.0 release builds; the raw serial transcripts are in
the repository
(\texttt{community/phase2/serial-frdm-*.log}) alongside the
consolidated \texttt{results-frdm.md}.

\subsection{Experimental Setup}
\label{sec:eval:setup}

The two targets follow Phase~1 (\cref{tab:setup2}): the
FRDM-MCXN947 board image with Zephyr shell and a 128~KB arena, and
the CI-oriented QEMU image with an 8~KB arena inside the emulator's
64~KB SRAM budget. Both link Zephyr v3.7.0~\cite{zephyr} with
\texttt{-Os} under Zephyr SDK 0.16.8. The model stage executes the
deterministic stub kernel on both targets in this phase; the eIQ
Neutron invoke path~\cite{nxp-neutron} remains future integration
work, so the latency numbers bracket the \emph{engine}, not the
accelerator. New relative to Phase~1, the board image routes the
DSP HAL to the PowerQuad engines (\cref{sec:processors:dsp}) ---
those numbers are real hardware, not stub.

\begin{table}[t]
\caption{Evaluation targets.}
\label{tab:setup2}
\centering
\renewcommand{\arraystretch}{1.15}
\begin{tabular}{@{}lll@{}}
\toprule
\textbf{Parameter} & \textbf{FRDM-MCXN947} & \textbf{QEMU} \\
\midrule
CPU              & Cortex-M33 @ 150 MHz   & Cortex-M3 (emulated) \\
FPU              & Hardware               & Soft-float \\
NPU backend      & Neutron (stub kernel)  & Software stub \\
DSP backend      & PowerQuad (hardware)   & Software kernels \\
SRAM (linker)    & 320 KB                 & 64 KB \\
Arena size       & 128 KB                 & 8 KB \\
Zephyr           & v3.7.0                 & v3.7.0 \\
Toolchain        & SDK 0.16.8, \texttt{-Os} & SDK 0.16.8, \texttt{-Os} \\
\bottomrule
\end{tabular}
\end{table}

\subsection{Build Footprint}
\label{sec:eval:size}

\Cref{tab:size2} reports the footprints of the two samples on both
targets from the v0.2.0 release builds, alongside the Phase~1
baselines. The FRDM images carry the Zephyr shell and the full
PowerQuad integration; the QEMU images build without the shell
(\texttt{hello\_inference}) and without PowerQuad (both), with the
logging subsystem enabled for \texttt{face\_detection}.

\begin{table}[t]
\caption{Build footprint (flash = text+data, RAM = data+bss; Phase
1 baselines from \cite{synapticos-p1}). FRDM Phase-2 rows include
shell and PowerQuad.}
\label{tab:size2}
\centering
\renewcommand{\arraystretch}{1.15}
\setlength{\tabcolsep}{4pt}
\begin{tabular}{@{}llrr@{}}
\toprule
\textbf{Build} & \textbf{Target} & \textbf{Flash} & \textbf{RAM} \\
\midrule
hello\_inference (P1)  & FRDM (shell) & 67.0 KB & 184.5 KB \\
hello\_inference (P2)  & FRDM (shell+PQ) & 87.7 KB & 201 KB \\
hello\_inference (P1)  & QEMU         & 24.1 KB & 27.8 KB \\
hello\_inference (P2)  & QEMU         & 27.9 KB & 30.4 KB \\
face\_detection (P2)   & FRDM (shell+PQ) & 88.8 KB & 203 KB \\
face\_detection (P2)   & QEMU         & 42.7 KB & 32.2 KB \\
\bottomrule
\end{tabular}
\end{table}

\begin{figure}[t]
\centering
\begin{tikzpicture}
\begin{axis}[
  width=\columnwidth, height=48mm,
  ybar, bar width=7.5mm,
  ymin=0, ymax=100,
  ylabel={flash (KB)},
  symbolic x coords={hello FRDM, hello QEMU, face FRDM, face QEMU},
  xtick={hello FRDM, hello QEMU, face FRDM, face QEMU},
  xticklabel style={font=\scriptsize, align=center},
  yticklabel style={font=\scriptsize},
  label style={font=\footnotesize},
  legend style={font=\scriptsize, at={(0.98,0.98)}, anchor=north east,
                draw=black!30},
  axis lines=left,
  enlarge x limits=0.16,
  nodes near coords, nodes near coords style={font=\tiny},
]
\addplot[fill=black!15, draw=black!60] coordinates {
  (hello FRDM,67.0) (hello QEMU,24.1)
};
\addplot[fill=blue!35, draw=black!60] coordinates {
  (hello FRDM,87.7) (hello QEMU,27.9) (face FRDM,88.8) (face QEMU,42.7)
};
\legend{Phase 1, Phase 2}
\end{axis}
\end{tikzpicture}
\caption{Flash footprint by build (v0.2.0, \texttt{-Os}). The
Phase-1-to-Phase-2 delta is +3.8~KB on the no-shell QEMU image (the
engine alone) and +20.7~KB on the shell-equipped FRDM image, which
additionally includes the PowerQuad routing, self-calibration, and
bench harness. \texttt{face\_detection} has no Phase~1 counterpart;
its FRDM image (shell and PowerQuad included) carries the complete
vision stack in 88.8~KB.}
\label{fig:footprint2}
\end{figure}
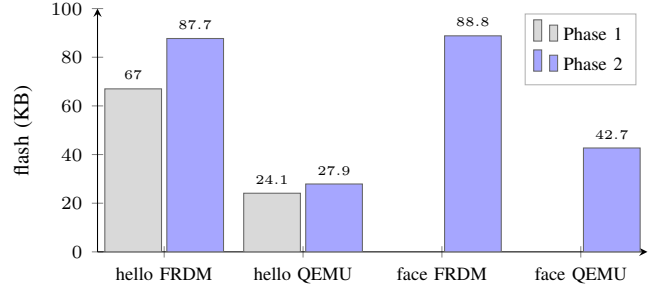

The Phase~1-to-Phase~2 delta on the QEMU image --- which contains
the engine and nothing else that changed --- is 3.8~KB of flash for
pipeline construction and execution, the job scheduler, all nine
processors, and the software FFT and matrix kernels; QEMU RAM grows
2.6~KB, dominated by the scheduler thread's 2~KB stack plus the
static job and pipeline tables. The shell-equipped FRDM image grows
20.7~KB of flash, which additionally buys the PowerQuad driver
routing, its boot-time self-calibration, and the
\texttt{syn dsp bench} harness; FRDM RAM grows 16.5~KB, of which
14~KB is static PowerQuad working memory (8~KB FFT staging buffers
plus 6~KB bench buffers). The complete \texttt{face\_detection}
vision stack --- pipeline, scheduler, processors, PowerQuad, and
shell --- fits in 88.8~KB of flash, within 1.1~KB of
\texttt{hello\_inference}. \Cref{fig:footprint2} visualises the
phase-over-phase growth.

\subsection{Engine Overhead on the Synchronous Path}
\label{sec:eval:latency}

The worst case for engine overhead is
\texttt{syn\_infer\_run\_sync()}, which pays for pipeline creation,
build, submission, the scheduler round-trip, result copy, and
pipeline destruction on every call (\cref{sec:sched:runsync}). We
measure it with the same workload as the Phase~1 latency experiment
--- the \texttt{test\_classify} model over a
$16{\times}16{\times}3$ INT8 input through the stub kernel --- so
the two paths are directly comparable. \Cref{lst:helloqemu} shows
the QEMU transcript.

\begin{lstlisting}[style=shellstyle,float=tp,caption={\texttt{hello\_inference}
via the scheduler path (QEMU, stub NPU, icount shift=6).},
label={lst:helloqemu},captionpos=b]
Pipeline 'run_sync' built: 1 stages, est. 4106 bytes
Inference completed in 1361 us
=== Inference Profile ===
  Total:       1010 us
  Preprocess:  4 us
  NPU:         1003 us
  Postprocess: 5 us
  Memory peak: 896 bytes
  NPU util:    99%
\end{lstlisting}

\begin{figure}[t]
\centering
\begin{tikzpicture}
\begin{axis}[
  width=0.81\columnwidth, height=46mm,
  xbar stacked, bar width=5mm,
  xmin=0, xmax=1500,
  xlabel={\textmu s (stub NPU on both targets)},
  symbolic y coords={{P1 direct (QEMU)}, {P2 run\_sync (QEMU)},
                     {P1 direct (FRDM)}, {P2 run\_sync (FRDM)}},
  ytick=data,
  yticklabel style={font=\scriptsize},
  xticklabel style={font=\scriptsize},
  label style={font=\footnotesize},
  legend style={font=\scriptsize, at={(0.5,-0.32)}, anchor=north,
                draw=black!30, /tikz/every even column/.append style={
                column sep=6pt}},
  legend columns=3,
  axis lines=left,
  enlarge y limits=0.18,
]
\addplot[fill=orange!45, draw=black!60] coordinates {
  (781,{P1 direct (QEMU)}) (1003,{P2 run\_sync (QEMU)})
  (1038,{P1 direct (FRDM)}) (1068,{P2 run\_sync (FRDM)})
};
\addplot[fill=blue!40, draw=black!60] coordinates {
  (0,{P1 direct (QEMU)}) (7,{P2 run\_sync (QEMU)})
  (0,{P1 direct (FRDM)}) (1,{P2 run\_sync (FRDM)})
};
\addplot[fill=black!20, draw=black!60] coordinates {
  (0,{P1 direct (QEMU)}) (351,{P2 run\_sync (QEMU)})
  (0,{P1 direct (FRDM)}) (61,{P2 run\_sync (FRDM)})
};
\legend{stub kernel, engine dispatch, run\_sync wrapper}
\end{axis}
\end{tikzpicture}
\caption{Wall-time decomposition of the synchronous convenience path
against the Phase~1 direct-HAL bracket, under QEMU (icount shift=6)
and on the FRDM-MCXN947. On the board, the 1{,}130~\textmu s wall
time splits into 1{,}068~\textmu s of stub model stage, 1~\textmu s
of engine dispatch inside the profiled window, and 61~\textmu s of
\texttt{run\_sync} wrapper (transient pipeline create/build/destroy,
submission, scheduler round-trip, wait, result copy) --- 92~\textmu s
total over the Phase~1 bracket. The QEMU wrapper share
(351~\textmu s) is inflated by the emulator's logging and
thread-switch costs. Stub-kernel baselines; not silicon throughput.}
\label{fig:latency2}
\end{figure}
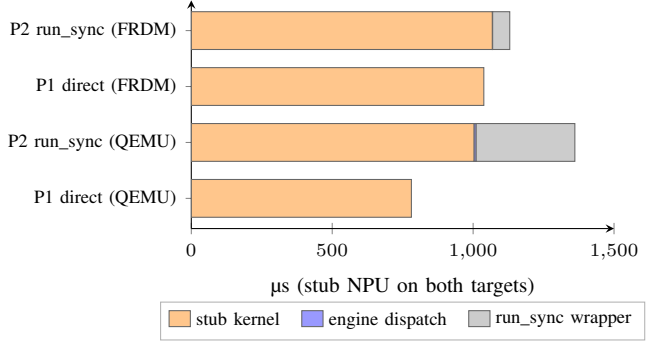

Three numbers matter, at three levels of the stack
(\cref{fig:latency2}):

\begin{itemize}
\item \textbf{7~\textmu s: dispatch overhead inside the profiled
window.} The profile's total-minus-NPU delta is the engine's cost
between the profiler marks --- stage-buffer allocation, stage
sequencing, and the profiler itself. On a 1~ms stub inference this
is the 99\% NPU-utilisation figure printed above.
\item \textbf{1{,}010~\textmu s: the profiled inference.} Almost
entirely the stub kernel's $O(n)$ pass, consistent with the Phase~1
direct-HAL measurements.
\item \textbf{1{,}361~\textmu s: wall time, submit to result.} The
580~\textmu s increase over the Phase~1 direct-HAL path
(781~\textmu s on the same emulator~\cite{synapticos-p1}) buys the
entire OS surface this paper describes, and --- because this is the
\texttt{run\_sync} convenience path --- includes transient pipeline
create, build (with its \texttt{LOG\_INF} line), destroy, and the
scheduler thread round-trip. Applications that build a pipeline once
and stream frames through it (\cref{sec:eval:facedet}) amortise all
of that away.
\end{itemize}

The QEMU figures are emulated-time baselines useful for isolating
engine costs deterministically and regression-testing them in CI;
the board tells the real story, and it is better. The same
\texttt{syn infer run test\_classify} workload on the FRDM-MCXN947
(\cref{lst:frdmshell}) measures \textbf{1{,}130~\textmu s} wall
against the Phase~1 direct-HAL bracket of 1{,}038~\textmu s on the
same silicon~\cite{synapticos-p1}: the \emph{entire} scheduler path
--- transient pipeline create and build, submission, dispatch,
wait, result copy-out, and destroy --- costs \textbf{92~\textmu s}
on hardware. Inside the profiled window the breakdown is
1{,}069~\textmu s total with 1{,}068~\textmu s in the stub model
stage, leaving roughly \textbf{1~\textmu s} of engine dispatch
(stage-buffer allocation, sequencing, profiler marks) per
inference; the boot-time run reports the same 1{,}069~\textmu s
total at 99\% NPU utilisation. The large QEMU-to-board gap in the
wrapper share (351 vs.\ 61~\textmu s) is consistent with the
emulator's inflated cost for the logging and thread-switch paths;
both figures bound the same code.

\subsection{Streaming: face\_detection}
\label{sec:eval:facedet}

The 30-frame face-detection run (\cref{sec:streaming}) exercises
the amortised path: one pipeline built once, 30 REALTIME jobs, one
ephemeral reset per frame. On the board it averages
\textbf{4{,}632~\textmu s per frame (215.8~FPS)}, against
31{,}134~\textmu s (32.1~FPS) under QEMU soft-float, detecting the
synthetic face in every frame on both targets (30 detections in 30
frames, one per frame after NMS --- the frame source is
deterministic, so the detection sequence is
target-independent).\footnote{On the board, early frames overran
the deferred-logging buffer (``35 messages dropped''); the summary
block and the last ten frame lines are complete, and the surviving
per-frame timings are uniform at 4{,}626--4{,}644~\textmu s. The
log path is not on the measured inference path.} Three
observations attach to the board number:

\begin{itemize}
\item \textbf{Where the time goes.} The last-frame profile reads
3{,}562~\textmu s preprocess, 1{,}040~\textmu s stub model stage,
1~\textmu s postprocess (4{,}602~\textmu s total, 22\% NPU
utilisation). The Cortex-M33's hardware FPU cuts the float-heavy
resize/normalize preprocessing roughly $8\times$ against QEMU
soft-float --- the bulk of the 6.7$\times$ overall frame-time
improvement --- and leaves preprocessing at 77\% of the frame. The
stub stage's deterministic $\sim$1~ms is not a real detector's
cost; once real Neutron kernels land, that 77\% preprocessing share
is the bottleneck to attack (\cref{sec:discussion:roadmap}).
\item \textbf{Throughput, not latency-hiding.} FPS through a
serialised scheduler means frames processed one at a time
(\cref{sec:sched:policy}).
\item \textbf{Constant footprint, confirmed on hardware.} The
board's arena telemetry over the full run reports 120 allocations,
30 ephemeral resets, a peak of \textbf{2{,}784~B} --- the 2{,}656~B
buffer plan of \cref{fig:pipeline} plus descriptor overhead ---
and occupancy ending at zero, identical on every frame: the
constant-footprint sawtooth of \cref{fig:sawtooth}, with zero
fragmentation by construction. (QEMU's integer-KB display reports
the same peak as ``2~KB''.)
\end{itemize}

\subsection{PowerQuad on Silicon: A Missed Target, Reported}
\label{sec:eval:pq}

The phase plan set an acceptance criterion of ``$\geq$10$\times$
over the software kernels'' for the PowerQuad DSP paths. The
measured result is below that target, and we report it as measured
rather than re-scoping the criterion --- the honest-baseline
methodology is only worth having if it also governs the numbers
that disappoint. \Cref{lst:pqbench} reproduces the on-board
benchmark verbatim.

\begin{lstlisting}[style=shellstyle,float=tp,caption={PowerQuad
vs.\ software DSP kernels on the FRDM-MCXN947 (\texttt{syn dsp
bench}, captured 2026-07-12; text verbatim from
\texttt{serial-frdm-shell.log}).},label={lst:pqbench},captionpos=b]
uart:~$ syn dsp bench
FFT f32 256 pts x16:
  soft: 86184 us (5386 us/op)
  hal:  15641 us (977 us/op)
  speedup: 5.51x
  max err: 976 ppm of peak
MatMul q15 16x16 x200:
  soft: 2948 us (14740 ns/op)
  hal:  1768 us (8840 ns/op)
  speedup: 1.66x
  max err: 1 LSB
\end{lstlisting}

Three facts frame the numbers. First, \textbf{the hardware works
and is verified}: the boot-time self-calibration pass
(\cref{sec:processors:dsp}) confirmed the PowerQuad CFFT's $1/N$
output-gain model on silicon and passed the Q15 known-answer and
saturation checks before either path was enabled, and the accuracy
figures above --- 976~ppm of peak for the FFT, consistent with the
$\sim$13-bit input scaling of the fixed-point transform engine, and
1~LSB for the Q15 matmul --- are within the expected envelopes.

Second, \textbf{the measurement is end-to-end by design}: the
per-operation times include the full HAL wrapper cost ---
float$\leftrightarrow$fixed staging loops on the FFT path, a
\texttt{PQ\_SetConfig} on every call, and the HAL mutex --- because
that is the cost a pipeline stage actually pays. A
transform-engine-only comparison would look better and mean less.

Third, \textbf{the shortfall has a legible structure}. The
256-point FFT, where the transform is large enough to amortise
per-call overhead, achieves 5.51$\times$; the $16{\times}16$
matmul, an 8.8-vs-14.7~\textmu s-per-op contest where staging and
configuration are a large fraction of each call, achieves only
1.66$\times$. The known paths to closing the gap --- persistent
PowerQuad configuration across calls, batched staging, and larger
transform and matrix sizes as real workloads provide them --- are
scoped to Phase~3 (\cref{sec:discussion:limits}). Until then, the
routing already pays for real pipelines (a 256-point MFCC frontend
spends $5.5\times$ less time per FFT) while falling short of the
plan's headline.

\subsection{Test Coverage}
\label{sec:eval:tests}

The suite grows from Phase~1's 61 cases in 10 suites to
\textbf{99 cases in 13 suites}, all passing on
\texttt{qemu\_cortex\_m3} via twister (\cref{tab:tests2}); the
99-case binary executes in 1.1~s of emulator time. Four suites are
new: pipeline construction and execution (8 cases), the job
scheduler (7), the FFT kernel (9), and the nine processors (15).
The Phase~1 placeholder \texttt{syn\_scheduler\_suite} (one
smoke-test case) is superseded by the real
\texttt{syn\_sched\_suite}, and two Phase~1 DSP cases flipped their
expectation from \texttt{-ENOTSUP} to success when FFT and Q15
matrix multiply became real (\cref{sec:processors:dsp}); no other
Phase~1 tests changed.

\begin{table}[t]
\caption{Phase 2 test suite (\texttt{qemu\_cortex\_m3}, captured
2026-07-11). Suites marked $\ast$ are new in Phase 2.}
\label{tab:tests2}
\centering
\renewcommand{\arraystretch}{1.12}
\begin{tabular}{@{}lrr@{}}
\toprule
\textbf{Suite} & \textbf{Cases} & \textbf{Pass} \\
\midrule
\texttt{syn\_mem\_suite}          & 18 & 18 \\
\texttt{syn\_process\_suite}$^\ast$ & 15 & 15 \\
\texttt{syn\_dsp\_suite}          & 11 & 11 \\
\texttt{syn\_dsp\_fft\_suite}$^\ast$ & 9 & 9 \\
\texttt{syn\_model\_suite}        &  9 &  9 \\
\texttt{syn\_pipeline\_suite}$^\ast$ & 8 & 8 \\
\texttt{syn\_sched\_suite}$^\ast$ &  7 &  7 \\
\texttt{syn\_npu\_suite}          &  6 &  6 \\
\texttt{syn\_dsp\_verify\_suite}  &  5 &  5 \\
\texttt{syn\_init\_suite}         &  4 &  4 \\
\texttt{syn\_mem\_bench\_suite}   &  3 &  3 \\
\texttt{syn\_mem\_regions\_suite} &  3 &  3 \\
\texttt{syn\_ipc\_suite}          &  1 &  1 \\
\midrule
\textbf{Total}                    & \textbf{99} & \textbf{99} \\
\bottomrule
\end{tabular}
\end{table}

The scheduler suite deserves specific mention: because ztest threads
on the emulator are cooperative, thread interleavings around the
completion protocol are \emph{deterministic}, and one of the seven
cases (\texttt{test\_priority\_order}) reproducibly caught a
callback/slot-reuse race in the first scheduler implementation on
every run. \Cref{sec:discussion:race} reconstructs the bug; we note
it here because it is the strongest evidence in this paper that the
CI-first methodology carries its weight --- a race that bit
deterministically in CI would have been an intermittent field bug
on preemptive hardware.

\subsection{Cross-Target Summary}
\label{sec:eval:summary}

\Cref{tab:frdm-summary} consolidates the headline measurements
across the two targets. Model-stage numbers are stub-NPU baselines
on both; the PowerQuad rows are real hardware.

\begin{table}[t]
\caption{Cross-target summary. Stub-NPU baselines except the
PowerQuad rows (real hardware); the phase plan's PowerQuad target
was $\geq$10$\times$.}
\label{tab:frdm-summary}
\centering
\footnotesize
\renewcommand{\arraystretch}{1.2}
\setlength{\tabcolsep}{3pt}
\begin{tabular}{@{}lrr@{}}
\toprule
\textbf{Metric} & \textbf{QEMU (stub)} & \textbf{FRDM} \\
\midrule
run\_sync wall time            & 1361 \textmu s & 1130 \textmu s \\
\quad P1 direct-HAL baseline   & 781 \textmu s  & 1038 \textmu s \\
\quad scheduler-path cost      & 580 \textmu s  & 92 \textmu s \\
Dispatch overhead (Total$-$NPU) & 7 \textmu s   & 1 \textmu s \\
face\_detection frame / FPS    & 31.1 ms / 32.1 & 4.63 ms / 215.8 \\
\quad preprocess share         & --- & 77\% \\
Arena peak per frame ($\to$ 0) & 2 KB (int.\ display) & 2784 B \\
PowerQuad FFT speedup (256-pt) & n/a (soft only) & 5.51$\times$ \\
PowerQuad matmul speedup ($16{\times}16$) & n/a (soft only)
  & 1.66$\times$ \\
\bottomrule
\end{tabular}
\end{table}

\section{Discussion}
\label{sec:discussion}

\subsection{Case Study: The Callback/Slot-Reuse Race}
\label{sec:discussion:race}

The completion-protocol ordering of \cref{sec:sched:completion}
exists because our first implementation got it wrong, and the manner
in which the test suite caught it is instructive beyond this
codebase.

\textbf{The bug.} The original scheduler loop completed a job in the
natural-seeming order: record the result, give the completion
semaphore, then fire the completion callback from the job slot.
Under ztest's cooperative threading, \texttt{k\_sem\_give()} on a
semaphore a cooperative waiter is pending on transfers control
\emph{immediately}: the waiter ran, consumed the result via
\texttt{syn\_infer\_get\_result()} (freeing the slot), and
resubmitted --- reinitialising the same slot for a new job --- all
before the scheduler thread resumed. When it did resume, it fired
the completion callback with the \emph{new} job's parameters: wrong
callback pointer, wrong user data, wrong job ID.

\textbf{The detection.} \texttt{test\_priority\_order} submits jobs
across the three priority classes and asserts on the execution order
its callbacks record. The corrupted callback context surfaced as an
impossible assertion: \texttt{exec\_order[0]} claimed a
\textsc{best-effort} job had run before a \textsc{realtime} job.
Because ztest threads are cooperative, the interleaving was not a
one-in-a-thousand window but the \emph{guaranteed} schedule: the
test failed on every run, bisected cleanly, and reproduced in
seconds under QEMU.

\textbf{The fix} is the protocol now described in
\cref{sec:sched:completion}: snapshot the callback, user data, job
ID, and output descriptor under the engine mutex; fire the callback
from the snapshot; give the semaphore last.

\textbf{The lesson} generalises. On preemptive hardware this race
window is a handful of instructions wide and priority-dependent ---
the classic intermittent field bug. A cooperative-threading test
harness inverts the economics: any race whose losing interleaving
the scheduler can express becomes a deterministic failure. We now
treat ``the completion path has a ztest case whose cooperative
schedule exercises the worst interleaving'' as an acceptance
criterion for scheduler changes, and we suggest the pattern to
anyone building dispatch machinery on Zephyr: the emulator plus
cooperative threads is not a weaker approximation of the board ---
for concurrency validation it is strictly stronger.

\subsection{Limitations}
\label{sec:discussion:limits}

As in the Phase~1 paper, we collect every known gap in one place.

\textbf{Stub NPU baseline.} Every latency and utilisation number in
\cref{sec:eval} that involves the model stage brackets the
deterministic stub kernel --- under emulation for the QEMU figures,
on real Cortex-M33 silicon for the FRDM figures. They isolate and
regression-pin the engine's overhead; they predict nothing about
Neutron silicon throughput. The eIQ Neutron SDK invoke path remains
the top integration item. (The PowerQuad measurements are the
exception: real hardware end to end.)

\textbf{Deadline dispatch and preemption are not implemented.}
\texttt{deadline\_us} and \texttt{preemptible} are accepted,
recorded, and inert. Dispatch is strict-priority, non-preemptive;
the worst-case priority inversion is one full pipeline execution
(\cref{sec:sched:policy}). Deadline-aware ordering and
layer-granularity preemption points are Phase~3, and the frozen API
already carries the parameters so applications written today need
no signature change.

\textbf{The PowerQuad speedup is below the plan target.} The phase
plan's acceptance criterion was $\geq$10$\times$ over the software
kernels; the board measures 5.51$\times$ (256-point FFT) and
1.66$\times$ ($16{\times}16$ Q15 matmul), wrapper costs included
(\cref{sec:eval:pq}). We record this as a missed criterion, not a
re-scoped one, and we consider doing so a feature of the
methodology rather than an embarrassment of the hardware: the
PowerQuad itself is routed, boot-time self-calibrated (gain model
and known-answer checks on silicon), and accurate to 976~ppm /
1~LSB --- what falls short is our wrapper, whose
float$\leftrightarrow$fixed staging loops, per-call
\texttt{PQ\_SetConfig}, and mutex acquisition dominate small
operations, and a $16{\times}16$ matrix is small. Persistent
PowerQuad configuration, batched staging, and larger
transform/matrix sizes are the known paths to closing the gap and
are Phase~3 backlog items; the criterion stays on the books until
one of them meets it.

\textbf{MFCC is not librosa-parity.} Non-overlapping frames, no
pre-emphasis, natural log, unnormalized DCT-II
(\cref{sec:processors:builtin}). Correct for models trained with
this frontend; a parity mode is future work.

\textbf{One model stage per pipeline; four pipelines.} Cascades
(detector feeding a classifier) currently compose at the application
level from multiple pipelines. Multi-model pipelines and a larger
pool are straightforward extensions the static-pool design admits.

\textbf{Results live in the ephemeral arena.} An output descriptor
is valid only until the next \texttt{syn\_mem\_reset\_ephemeral()}.
This is the contract that buys constant-footprint streaming
(\cref{sec:streaming}), but it is a sharper lifetime than
heap-returned results; \texttt{run\_sync} therefore copies into a
caller buffer when one is provided.

\textbf{Single-slot profiler.} The profiler retains the last
completed inference only; serialised execution keeps records
coherent, but a high-rate client wanting per-job attribution must
read between completions (\cref{sec:profiling}).

\subsection{Roadmap}
\label{sec:discussion:roadmap}

Phase~3 carries the scheduling story forward: deadline-aware
dispatch over the already-recorded \texttt{deadline\_us},
layer-granularity preemption points (making \texttt{preemptible}
meaningful), TFLite Micro~\cite{tflm} as a production model stage,
the real Neutron invoke path, and the MCXN947's second Cortex-M33
core --- at which point the serialised-execution simplification of
\cref{sec:sched:policy} is deliberately broken and the completion
protocol of \cref{sec:sched:completion} earns its keep under true
concurrency. On the DSP side, Phase~3 carries the PowerQuad wrapper
work needed to revisit the missed speedup target --- persistent
configuration, batched staging --- which also attacks the 77\%
preprocessing share the board's face-detection profile exposed
(\cref{sec:eval:facedet}). Camera and LCD bring-up (deferred from
this phase) turn the face-detection sample into an end-to-end
hardware demo. The later-phase items (OTA model updates, fault
recovery, additional vendor backends) are unchanged from the
Phase~1 roadmap~\cite{synapticos-p1}.

\section{Conclusion}
\label{sec:conclusion}

We presented the Phase~2 inference engine of SynapticOS, which
promotes the inference pipeline from an application-code convention
to an operating-system object. A pipeline is drawn from a static
pool, validated at construction against a canonical
pre$^{*}$-model-post$^{*}$ grammar, given exact-fit intermediate
buffers planned from stage configuration and runtime tensor geometry
(with a bounded $4\times$ fallback for stages the engine cannot see
inside), and executed by a priority job scheduler --- three classes,
FIFO within class, per-job completion semaphores, defined
cancellation, bounded job table, no heap anywhere on the path.
Because every intermediate lives in the ephemeral arena region and
streaming applications reset that region per frame, memory behaviour
is a constant-footprint sawtooth with zero fragmentation across
unbounded frame counts. The Phase~1 profiler is now driven at stage
boundaries, closing the known instrumentation gap live on the board.

Measured on the FRDM-MCXN947 through the deterministic stub NPU ---
engine baselines, not silicon claims --- the entire scheduler path
costs 92~\textmu s of wall time over the Phase~1 direct-HAL bracket
(1{,}130 vs.\ 1{,}038~\textmu s), with roughly 1~\textmu s of
dispatch inside the profiled window; the 30-frame face-detection
pipeline averages 4.63~ms per frame (215.8~FPS, 6.7$\times$ the
QEMU soft-float figure) at a constant 2{,}784-byte arena peak that
returns to zero after every frame. The PowerQuad DSP is routed and
boot-time self-calibrated; its measured end-to-end speedups ---
5.51$\times$ for the 256-point FFT, 1.66$\times$ for the
$16{\times}16$ Q15 matmul, wrapper costs included --- fall short of
the phase plan's $\geq$10$\times$ target, and we report the miss
and its structure rather than re-scope it. The engine adds 3.8~KB
of flash to the QEMU image and 20.7~KB to the shell-equipped FRDM
image including the PowerQuad integration. Test coverage grows from
61 cases in 10 suites to 99 cases in 13, at a 100\% pass rate ---
and the suite's cooperative-scheduling determinism caught a
completion-protocol race that would have shipped as an intermittent
field bug, a methodological result we consider as valuable as the
engine itself.

SynapticOS v0.2.0, the test suite, the QEMU and FRDM measurement
artifacts (including the raw serial transcripts behind every board
number in this paper), and the LaTeX sources of this paper are
released under Apache~2.0 at
\url{https://github.com/Dimitrios-Kafetzis/SynapticOS}.

\bibliographystyle{IEEEtran}
\begingroup
\emergencystretch=1em
\bibliography{refs}
\endgroup

\end{document}